\documentclass[%
	reprint,
	amsmath,amssymb,
	aps,
	prd,
]{revtex4-1}

\usepackage{gensymb}
\usepackage{graphicx}% Include figure files
\usepackage{multirow}% Include figure files
\usepackage{dcolumn}% Align table columns on decimal point
\usepackage{bm}% bold math
\begin{document}

\preprint{APS/123-QED}

\title{Geoneutrinos at Jinping: Flux prediction and oscillation analysis}% Force line breaks with \\

\author{Linyan Wan}
\email{wanly13@mails.tsinghua.edu.cn}
\author{Ghulam Hussain}
\email{gu-lm11@mails.tsinghua.edu.cn}
\author{Zhe Wang}%
\author{Shaomin Chen}%
\affiliation{%
	Department of Engineering Physics, Tsinghua University.
}%

\date{\today}% It is always \today, today,

\begin{abstract}
	Geoneutrinos are electron antineutrinos ($\bar\nu_e$) generated by the beta-decays of radionuclides naturally occurring inside the Earth, in particular $^{238}$U, $^{232}$Th, and $^{40}$K.
	Measurement of these neutrinos provides powerful constraints on the radiogenic heat of the Earth and tests on the Earth models. %radioactive heat
	Since the prediction of $\bar\nu_e$'s in geoneutrino flux is subject to neutrino oscillation effects, we performed a calculation including detailed oscillation analysis in the propagation of geoneutrinos and reactor neutrinos generated around the Earth.
	The expected geoneutrino signal, the reactor neutrino background rates and the systematic error budget are provided for a proposed 3-kiloton neutrino detector at the Jinping underground lab in Sichuan, China. 
	In addition, we evaluated sensitivities for the geoneutrino flux, Th/U ratio and power of a possible fission reactor in the interior of the Earth. 

	\begin{description}
%		\item[Key words]Geo-neutrinos; Neutrino oscillation; Jinping; Th/U ratio; BSE model test
		\item[DOI]
		\item[PACS numbers]14.60.Lm, 14.60.Pq, 91.35.-x, 91.67.Qr
	\end{description}
\end{abstract}
\maketitle

\section{Introduction}

\subsection{Energy Budget of the Earth}

The energy budget of the Earth is an important quantity in many fundamental geological questions, as it touches on the composition of the Earth, chemical layering in the mantle, the power source of mantle convection, plate tectonics, and the geodynamo, which generates the magnetosphere that protects the planet from cosmic radiation~\cite{McDonough:2014}.
The Earth surface heat flow is currently estimated to be $46\pm3$ TW~\cite{Jaupart:2007, JH:2010}.
The driving power comes mainly from the radiogenic energy of the heat producing elements (HPE) potassium, thorium and uranium, and the initial inheritance of primordial energy that resulted from the accretion of the planet and the gravitational differentiation of metal sinking to the center of the Earth.

There are several bulk silicate Earth (BSE) models estimating the chemical composition of the primitive mantle, categorized in three classes of distinct predictions for BSE radiogenic heat (Q):
a) the low-Q models have the lowest value ($11\pm2$ TW)~\cite{Javoy:2010};
b) the medium-Q models predict a median amount ($20\pm4$ TW)~\cite{ArevaloJr:2010};
c) the high-Q models have the highest prediction ($33\pm3$ TW)~\cite{turcotte:1982}.

The Earth's magnetic field is continuously consuming energy to power itself and its long-term variation.
A natural self-sustaining fission reactor mechanism at the center of Earth was proposed by J.M. Herndon in the 1990s as one of the possible explanations of this phenomenon~\cite{Herndon:2001}.

\subsection{Geoneutrinos}

One of the best ways to experimentally measure the radiogenic power is to measure the amount of neutrinos coming from the interior of the Earth.
Geoneutrinos are electron antineutrinos generated from radioactive decay chains inside the Earth, with typical energies below 3.3 MeV.
Because of their extremely low cross section with matter, geoneutrinos act as messengers with information on the HPE distribution inside the Earth, thus providing an insight into radiogenic Earth models.

Another contribution to the geoneutrino flux may be the electron antineutrinos coming from the hypothesized Earth core fission processes~\cite{Raghavan:2002}, which could shed light on the geology inside the Earth core.
Energy spectra of such Earth core fission neutrinos are different from the geoneutrinos from radioactive decay chains and extend up to more than 8 MeV.

A precisely measured geoneutrino rate and energy spectrum could allow an identification of mantle geoneutrinos~\cite{Mantle:2012}, leading to a new determination of the radiogenic power in the Earth's thermal energy budget, a discrimination between different BSE predictions, and a conclusive confirmation of the Earth core fission hypothesis.

In following context, we discuss mainly geoneutrinos from radioactive decay chains and refer to them as geoneutrinos.
A discussion on Earth core fission neutrinos is included in Sec. VII.

\subsection{Geoneutrino Experiments}

The study of geology with the elusive geoneutrinos~\cite{McDonough:2014} only became practical recently with the advent of neutrino detectors~\cite{KL:2005, KL:2011, KL:2013, BX:2010, BX:2013, BX:2015}.
KamLAND~\cite{KL:2013} and Borexino~\cite{BX:2015} performed experimental studies with large volume liquid scintillator detectors and reported positive observations of geoneutrinos.
Both studies disfavor the Earth model that driving power comes only from radiogenic energy, yet cannot distinguish among the predictions due to the detection uncertainty contributed mainly by the low statistics, the backgrounds, and the uncertainty of HPE distribution in the Earth.
In addition, the ratio of uranium and thorium in geoneutrinos are still limited by the low statistics and the backgrounds.
Additional experimental approaches are needed.

Future experiments like SNO+~\cite{SNO+:2015}, JUNO~\cite{JUNO:2015, JUNO:PI}, HANOHANO~\cite{HANOHANO:2012}, and Jinping~\cite{JinpingNE:2016} will push forward the detection of geoneutrinos.

\subsection{Geoneutrinos with Jinping Neutrino Experiment}

The Jinping Neutrino Experiment (Jinping) is a proposed neutrino observatory for low-energy neutrino physics in the China JinPing Laboratory (CJPL, 28.15323$\degree$N, 101.7114$\degree$E), an ideal site to do low background experiments.
The experimental site is located in Jinping Mountain, Sichuan Province, China, at least 950 km away from all the nuclear power plants in operation and under construction.
The detector is designed to use a liquid scintillator or slow liquid scintillator, with a fiducial mass of 3 kilotons for inverse beta decay (IBD) events.
Initial sensitivity studies for the Jinping detector based on assessments of the site and potential detector designs have been conducted~\cite{JinpingNE:2016}. 

In this paper, the geoneutrino spectra are discussed in Sec. II and the calculation of the predicted geoneutrino signals with oscillation analysis is presented in Jinping in Sec. III.
The detection method and background analysis are included in Secs. IV and V.
The evaluation of geoneutrino measurements in Jinping is presented in Sec. VI.
Additional discussion on Earth core fission neutrinos is included in Sec. VII.
%Compared with different experiments and theoretical predictions, a determination of the mantle neutrino ratio in the geoneutrino flux can be extrapolated.

%----------------------------------------------------------------------------------------
%  METHODS
%----------------------------------------------------------------------------------------

\section{Antineutrino Intensity Energy Spectra}

Radioisotopes that are abundant in the Earth are categorized into three major types; isotopes in the $^{232}\rm{Th}$  $(\tau_{\frac{1}{2}} = 14.0  \times 10^9 \rm {\,year})$ decay chain, isotopes in the $^{238}\rm{U}$ $(\tau_{\frac{1}{2}} = 4.47  \times 10^9 \rm {\,year})$ decay chain,  and $^{40}\rm{K}$ $(\tau_{\frac{1}{2}} = 1.28  \times 10^9 \rm {\,year})$:
\begin{equation}
	\begin{aligned}
		&	^{238}\rm{U}\rightarrow ^{206}Pb + 8 \alpha + 6e^- + 6{\bar \nu_e} + 51.698~MeV,\\
		&	^{232}\rm{Th}\rightarrow ^{207}Pb + 7 \alpha + 4e^- + 4{\bar \nu_e} + 46.402~MeV,\\
		&	^{235}\rm{U}\rightarrow ^{208}Pb + 6 \alpha + 4e^- + 4{\bar \nu_e} + 42.652~MeV,\\
		&	^{40}\rm{K}\rightarrow ^{40}Ca + e^-  + 4{\bar \nu_e} +  1.311~MeV~~~(89.3\%),\\
		&	^{40}\rm{K} + e^- \rightarrow ^{40}Ar   + \nu_e + 1.505~MeV~~~(10.7\%).
	\end{aligned}
\end{equation}
Except for the K-shell electron capture of $^{40}$K, all the other $\beta$ decays produce $\bar{\nu}_e$'s, comprising the geoneutrinos.
It is noted that only those from $^{232}$Th and $^{238}$U decay chains with energy above IBD threshold of 1.8 MeV can be detected.
In the estimation of the overall antineutrino intensity energy spectrum of each decay series, the shapes and rates of all single decays have to be incorporated: comprehensive calculations are needed to take into account 82 individual branches in $^{238}\rm{U}$ and 70 individual branches in  $^{232}\rm{Th}$. 
The only contributions to the geoneutrino signal detectable via IBD (see Sec. IV) are from $^{214}\rm{Bi}$ and $^{234}\rm{Pa}$ in the $^{238}\rm{U}$ series and $^{212}\rm{Bi}$ and $^{228}\rm{Ac}$ in the $^{232}\rm{Th}$ series~\cite{Ludhova:2013hna}.

The energy spectrum of each beta decay with maximum electron energy $E_{max}$ is followed by the allowed decay formula~\cite{Enomoto:2005},
\begin{equation}
	\begin{aligned}
		dN(E_e) = &\frac{G^2_{F}|M|^2}{2\pi^3 \hbar^7 c^5} F(Z, E_e)(E_{max}-E_e)^2 \\
		&\times\sqrt{E^2_e-m^2_ec^4}E_edE_e,
	\end{aligned}
\end{equation}
where $F(Z,E_e)$ is the Fermi function for the effect of the electrical field of the nucleus. For each branch, the energy of  the antineutrino $E_{\bar\nu_{e}}$ is given by
\begin{equation}
E_{\bar\nu_e} = E_{max} - E_e.
\end{equation}

\begin{figure}[htbp] %  figure placement: here, top, bottom, or page
\centering
\includegraphics[width=1.\columnwidth]{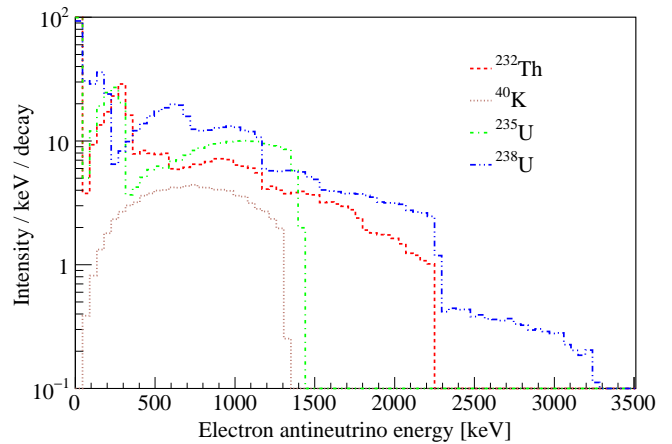}
	\caption{The geoneutrino energy spectra produced by HPEs simulated with Geant4.
	 The $\bar\nu_{e}$'s are produced in the decays of  $^{238}\rm{U}$, $^{232}\rm{Th}$, $^{40}\rm{K}$ and  $^{235}\rm{U}$.  $\nu_e$ from $^{40}\rm{K chain}$ are not shown. }
\label{fig:nue_inetensity}
\end{figure}

Adding up all the antineutrino intensity spectra from the individual beta decays gives a total spectrum as shown in Fig.~\ref{fig:nue_inetensity}. The individual decay chain spectra are also shown. All spectra were generated with Geant4~\cite{Geant4:2002}. The total geoneutrino spectrum has a maximum end point at 3.3 MeV and the contribution from different nuclides can be identified according to their end points; e.g., geoneutrinos with E $\textgreater$2.25 MeV are contributed by only the $\rm^{238}U$ series. %According to geochemical knowledge, $^{232}\rm{Th}$ is more abundant in nature than $^{238}\rm{U}$, and their mass fraction in the Earth is approximated to be m($^{232}$Th)/m($^{238}$U) = 3.9 as an average over BSE~\cite{Huang:2013}.

It is noted that there is a few percent difference around 1 MeV for $^{238}$U between the present Geant4 and S. Enomoto's calculation~\cite{Enomoto:2005}. Both the $^{235}\rm{U}$ and $^{40}\rm{K}$  geoneutrinos are below the IBD threshold of 1.8 MeV, as shown in Fig.~\ref{fig:nue_inetensity}, so they cannot be detected by the proposed techniques.

\section{Geo $\bar\nu_e$ Flux Calculation}

The geoneutrino energy spectrum $\phi(E)$ at the Jinping site emitted by an HPE is calculated by the integral of a grid-calculated geoneutrino flux in Earth propagating to Jinping with oscillation,
\begin{equation}
	\begin{aligned}
		\phi_i(E)dE=&\frac{X_i\lambda_i N_A}{\mu_i}n_\nu(i) \\
		&\times\int\frac{A_i(\vec r)\rho(\vec r)}{4\pi L^2}P_{ee}(E,L)f_i(E)d\vec rdE,
	\end{aligned}
	\label{Eq:fluxE}
\end{equation}
where $X$ represents the natural isotopic mole fraction of isotope $i$, $\lambda$ is the decay constant for $i$, $N_A$ is Avogadro's constant, $\mu$ is the standard atomic molar mass for $i$, and $n_\nu$ is the number of $\bar\nu_e$'s emitted per decay for $i$; $A(\vec r)$ and $\rho(\vec r)$ are respectively the locally variant Earth model parameter of abundance for $i$ and density; $L$ is the linear distance to the Jinping site; $P_{ee}$ is the neutrino survival probability in the framework of three generations of neutrinos, and $f(E)$ is the normalized electron antineutrino energy spectrum for $i$.

The total flux $\phi_i$ for HPE $i$ is obtained by integrating over the energy,
\begin{equation}
	\begin{aligned}
		\phi_i=&\int\phi_i(E)dE\\
		=&\frac{X_i\lambda_i N_A}{\mu_i}n_\nu(i) \langle P^i_{ee}\rangle\int\frac{A_i(\vec r)\rho(\vec r)}{4\pi L^2}d\vec r,
%	R=\frac{X\lambda N_A}{\mu}n_\nu(i) C \langle P_{ee}\rangle_r \iiint\frac{A(\vec r')\rho(\vec r')}{4\pi|\vec r'-\vec r|^2}d\vec r',
	\end{aligned}
	\label{Eq:flux}
\end{equation}
where $\langle P^i_{ee}\rangle$ is the electron antineutrino survival probability averaged over the energy spectrum and the geological distribution of isotope $i$ (see Sec. III. A).

\subsection{Earth Model}

A $1\degree\times1\degree$ topological map of the density $\rho(\vec r)$ in the Earth crust is used in Eqs. (\ref{Eq:fluxE}) and (\ref{Eq:flux}), and was obtained from CRUST1.0~\cite{CRUST:2014}. 
The assumption employed for the mantle is from Huang \textit{et al}~\cite{Huang:2013}.
For the computation of flux, a $1\degree\times1\degree$ tile is further divided into subtiles to obtain the propagation distance $L$.
The abundance of HPEs $A_i(\vec r)$ in geological layers and the intrinsic radioactive element properties are taken from Ref.~\cite{Sramek:2016}, assuming the medium-Q BSE model.
The abundance is assumed to be uniform in every layer.
The energy spectra of HPEs are obtained as in Sec. II.

According to Ref.~\cite{Sramek:2016}, the uncertainty on the geoneutrino flux prediction introduced by this Earth model is $^{+12.6\%}_{-12.3\%}$, while for crustal geoneutrinos, this uncertainty is $\pm$15.0\%.

\subsection{Oscillation Analysis}

\subsubsection{Vacuum Oscillation}

The survival probability of an electron antineutrino with energy $E$ propagating over a baseline $L$ can be written as
\begin{equation}
	P_{ee}(E,L)=|\sum_ie^{-\tilde M_{i,1}}\times U_m(0,i)^\dagger\times U_m(i,0)|^2,
\end{equation}
where $\tilde M_{i,1}\approx2.534\times\Delta M_{i1}L/E$, with $L$ in km and $E$ in GeV, and $\Delta M_{ij}$ is the neutrino mass difference between generation $i$ and $j$.
$U_m$ is the eigenmatrix of neutrino mass mixing matrix $A=U\times M\times U^\dagger$, where $M$ is the neutrino mass matrix, $M_{ij}=\delta_{ij}\times\Delta M_{ij}$, and $U$ is the neutrino oscillation matrix,
\begin{equation}
	\begin{aligned}
		U=&\left(\begin{array}{ccc} 
		1 & & \\
		& c_{23} & s_{23}\\
	 -s_{23} & c_{23} \end{array}\right)\times
		\left(\begin{array}{ccc}
			c_{13} & & s_{13}e^{-i\delta_{cp}}\\
			& 1 & \\
		-s_{13}e^{-i\delta_{cp}} & & c_{13} \end{array}\right)\\
		&\times\left(\begin{array}{ccc}
			c_{12} & s_{12} & \\
			-s_{12} & c_{12} & \\
		& & 1 \end{array}\right).
	\end{aligned}
\end{equation}
The central values and uncertainties of oscillation parameters $\theta_{ij}$ and $\Delta M_{ij}$ are taken from Ref.~\cite{PDG:2016}. 
The neutrino mass hierarchy is assumed to be inverted hierarchy.

\begin{figure}[!h]
	\centering
	\includegraphics[width=1.\columnwidth]{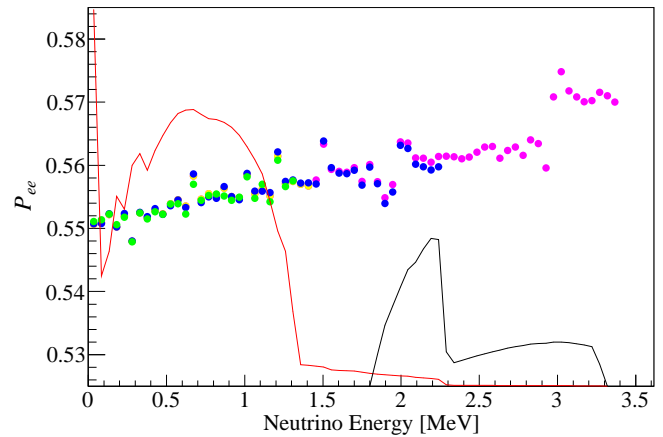}
	\caption[$\bar\nu_e$ survival probability]{Geoneutrino survival probability averaged over the HPE distributions in the Earth. Different colored points represent $P_{ee}$ for different HPEs (magenta for U, blue for Th, and green for K). Solid lines represent the total geoneutrino flux (red) and the IBD events (black) in Jinping in arbitrary units.}
	\label{fig:Pee}
\end{figure}

The average survival probability of geoneutrinos given in Eq.~(\ref{Eq:flux}) can be calculated as
\begin{equation}
	\begin{aligned}
		\langle P^i_{ee}\rangle&=\int P^i_{ee}(E)f_i(E)dE,\\
		P^i_{ee}(E)&=\frac{\int P_{ee}(E, L)\cdot A_i(\vec r)\rho(\vec r)/(4\pi L^2)d\vec r}{\int A_i(\vec r)\rho(\vec r)/(4\pi L^2)d\vec r}.
	\end{aligned}
\end{equation}
The average survival probability $P^i_{ee}(E)$ for HPEs is shown in Fig. \ref{fig:Pee}.

Table \ref{tab:Pee} lists for HPEs the $\langle P^i_{ee}\rangle$ and $\langle P^i_{ee}\rangle^{'}$, which is for the effective energy spectrum, i.e., it is weighted by the IBD cross section (see Sec. IV).
\begin{table}[h]
	\begin{center}
		\begin{tabular}[c]{lccccc} \hline\hline
			& $^{40}$K & $^{232}$Th & $^{235}$U & $^{238}$U & Total\\\hline
			$\langle P_{ee}\rangle$ & 0.554 & 0.553 & 0.553 & 0.554 & 0.553\\
			$\langle P_{ee}\rangle^{'}$ & 0     & 0.560 & 0     & 0.563 & 0.562\\\hline\hline
		\end{tabular}
		\caption{The average and effective (IBD weighted) average survival probabilities for geoneutrinos in Jinping.}
		\label{tab:Pee}
	\end{center}
\end{table}

\subsubsection{Uncertainty Introduced by Oscillation Parameters}

It should be noted that the uncertainties on the neutrino oscillation parameters will propagate into the uncertainty on the flux of geoneutrinos.
This uncertainty, especially the uncertainty in the crustal geoneutrino flux prediction, is essential for the test of BSE models and determination in the Earth energy budget.
Table \ref{tab:Uncer} lists all the uncertainties on the parameters from Ref.~\cite{PDG:2016} and their impacts on the flux prediction of geoneutrinos.
$\Delta M_{ij}$'s are in eV.
\begin{table}[h]
	\begin{center}
		\begin{tabular}[c]{ccccc} \hline\hline
			& & & Flux & Crustal \\
			Parameter & Value & Uncertainty & uncertainty & uncertainty\\\hline
			\multirow{ 2}{*}{$\theta_{12}$} & \multirow{ 2}{*}{0.584}   & +2.6\% & +1.8\%  & +1.8\% \\
								 & & $-$2.4\% & $-$1.7\% & $-$1.7\%\\
			\multirow{ 2}{*}{$\theta_{13}$}   & \multirow{ 2}{*}{0.149} & +2.7\% & \multirow{ 2}{*}{$\pm$0.2\%} & \multirow{ 2}{*}{$\pm$0.2\%} \\
								 & 		& $-$2.8\% & &  \\
			$\theta_{23}$   & 0.785 & $\pm$6.4\% & $\pm$0.0\% & $\pm$0.0\% \\
			$\Delta M_{21}$ & 7.53$\times 10^{-5}$ & $\pm$2.4\% & $\pm$0.1\% & $\pm$0.1\% \\
			$\Delta M_{32}$ & 2.51$\times 10^{-3}$ & $\pm$2.4\% & $\pm$0.0\% & $\pm$0.0\% \\
			$\delta$        & 0 & $\pm$1.5 & $\pm$0.0\% & $\pm$0.0\% \\
			MH        & IH & NH & $\pm$0.0\% & $\pm$0.0\% \\\hline\hline
		\end{tabular}
		\caption{The values of neutrino oscillation parameters, their relative uncertainties (parameter boundaries for $\delta$ and mass hierarchy MH) and the corresponding uncertainties on the flux of geoneutrinos.}
		\label{tab:Uncer}
	\end{center}
\end{table}
The uncertainty introduced by neutrino oscillation parameters is $^{+1.8\%}_{-1.7\%}$, which is smaller than the present $^{+12.6\%}_{-12.3\%}$ uncertainty from the Earth model.
However, the uncertainty in the present model simply scales with the lithospheric flux magnitude and is very likely to be further reduced with a new calculation.
In this case, a better measurement of $\theta_{12}$ would be beneficial. 

\begin{figure}[!h]
	\centering
	\includegraphics[width=1.\columnwidth]{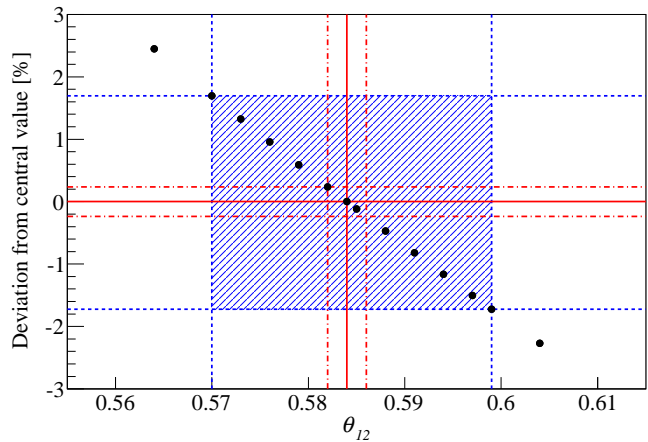}
	\caption[$\bar\nu_e$ survival probability]{The relative uncertainty on the geoneutrino survival probability vs the neutrino mixing angle $\theta_{12}$. The red solid line represents the central value, the blue dotted lines are for the present 1$\sigma$ region, and the red dotted dashed line represents the predicted uncertainty from JUNO~\cite{JUNO:PI}.}
	\label{fig:t12}
\end{figure}

With future solar and reactor neutrino experiments, a much more precise measurement of $\theta_{12}$ is expected.
Figure \ref{fig:t12} shows the uncertainty of $\theta_{12}$ propagated into the geoneutrino prediction.
The red solid line represents the central value $\theta_{12}=0.584$, while the blue dotted lines are for the present 1$\sigma$ uncertainty region, yielding $^{+1.8\%}_{-1.7\%}$ uncertainty in geoneutrino event rate prediction, and the red dotted dashed line represents the predicted uncertainty (systematic only) from JUNO~\cite{JUNO:PI}; the $\pm$0.3\% uncertainty predicted for $\theta_{12}$ will improve the corresponding uncertainty in the geoneutrino event rate prediction to $\pm$0.2\%.

\subsubsection{MSW Oscillation}

The three-generation Mikheyev-Smirnov-Wolfenstein (MSW) oscillation effect~\cite{Bahcall:1997} was also studied for geoneutrinos. 
The neutrino mixing matrix for the MSW oscillation changes from vacuum oscillation as
\begin{equation}
	A'_{11}=A_{11}-V
\end{equation}
for antineutrinos, where $V$ is the chemical potential,
\begin{equation}
	V=2\sqrt2G_Fn_eE_\nu,
\end{equation}
$G_F$ is the Fermi constant, $n_e$ is the electron density, and $E_\nu$ is the neutrino energy.
For simplicity, the electron density on Earth is estimated by the Preliminary Reference Earth Model with seven spherical layers~\cite{Giunti:1998}.

With the above assumption and similar analysis procedures as in the case of vacuum oscillation, the prediction given by the MSW oscillation has a $+$0.3\% deviation from the prediction of geoneutrino flux based on vacuum oscillation.

\section{Inverse Beta Decay Detection}

In principle, the produced electron antineutrinos could be detected via either the elastic scattering process or the IBD reaction, the former of which has a relatively low cross section and the signal signature overlays with a solar neutrino background.
In this paper, only the IBD reaction is discussed.

The IBD reaction has the signature of a prompt positron signal and a delayed neutron capture gamma correlated in both time and space~\cite{Vogel:1999},
\begin{equation}
	\begin{aligned}
		\bar{\nu}_e + p &\rightarrow e^+ + n,\\
		n + H &\rightarrow d + \gamma~(2.2 \mbox{MeV}),
	\end{aligned}
\end{equation}
with an energy threshold of 1.806 MeV. Therefore, the detectable signals are composed of the neutrinos from $^{238}$U and $^{232}$Th decays only. 
Also, the difference in the geoneutrino energy spectrum discussed in Sec. II does not influence the detection result with the IBD detection method.
The prompt positron will decelerate and annihilate with an electron, yielding gammas.
The energy deposited in the detector by the deceleration and annihilation of the positron, called visible energy, can be approximately calculated by $E_{vis}=E_{\bar\nu_e}-0.784$ MeV.
Here, $E_{vis}$ is the visible energy of the positron, and $E_{\bar\nu_e}$ is the initial neutrino energy. 
The tiny neutron recoil energy is neglected.
The neutron is detected via neutron capture on hydrogen or another nuclide.
For a liquid scintillator or slow liquid scintillator~\cite{Li:2015}, the neutron capture is on hydrogen, which generates a single gamma of 2.2 MeV.
A typical target proton number is $7.2\times10^{31}/$ kiloton for a slow scintillator.

To avoid redundant scaling between different target masses and live times, the unit Terrestrial Neutrino Unit (TNU) is introduced for the electron antineutrino events detected via the IBD reaction in the geoneutrino calculation.
1 TNU = 1 event / 10$^{32}$ protons / 1 year, assuming 100\% detection efficiency.
The IBD event rates in Jinping are calculated as
\begin{equation}
	R(E_{\nu}) = \phi(E_{\nu}) \times \sigma(E_{\nu}),
\end{equation}
where $\sigma(E_{\nu})$ is the IBD cross section.

\section{Backgrounds}

\subsection{Reactor Neutrino Background}

Reactor electron antineutrinos form an irreducible background to the detection of geoneutrinos.
With the same signal signature, this background can only be reduced by placing the detector far away from nuclear power plants, as the flux decreases by $1/L^2$.
The location of Jinping is at least 950 km away from nuclear power plants, making it the best site for a geoneutrino experiment among all the existing experimental sites, in terms of the signal-background ratio.

Reactor antineutrinos are from the beta decays of four main fissile nuclei $\mathrm{{}^{235}U}$, $\mathrm{{}^{238}U}$, $\mathrm{{}^{239}Pu}$, and $\mathrm{{}^{241}Pu}$. The differential $\bar{\nu}_e$ flux, $\phi(E_{\nu})$, for a reactor is estimated as
\begin{equation}
	\phi(E_{\nu})=\frac{W_{th}LF}{\sum_{i}f_{i}e_{i}}\sum_{i}f_{i} S_{i}(E_{\nu}),
	\label{equ:Neutrino_Flux}
\end{equation}
where $i$ sums over the four isotopes and $W_{th}$ is the thermal power of a reactor which can be found in International Atomic Energy Agency (IAEA)~\cite{IAEA:2015, Baldoncini:2015}.
$LF$ is the load factor~\cite{IAEA:2015}, taken as 1.0 uniformly in this study.
$f_i$ ($\sum_{i}f_i=1$) is the fission fraction of each isotope,
$e_i$ is the average energy released per fission of each isotope taken from Ref.~\cite{Ma:2012}, and $S_{i}(E_{\nu})$ is the antineutrino spectrum per fission of each isotope~\cite{Huber:2011, Mueller:2011}.
A set of typical fission fractions, $f_i$, and the average energy released
per fission, $e_i$, are listed in Table~\ref{tab:EnergyReleased}.
%The spectrum of each isotope, $S_{i}(E_{\nu})$, and their sum are shown in Fig.~\ref{fig:component}.

\begin{table}[h]
	\begin{center}
		\vspace*{0cm}
		\begin{tabular}[c]{ccc} \hline\hline
			Isotope               & $f_i$& $e_i$ [MeV/fission]   \\\hline
			$\mathrm{{}^{235}U}$  & 0.58 & $202.36\pm0.26$     \\
			$\mathrm{{}^{238}U}$  & 0.07 & $205.99\pm0.52$     \\
			$\mathrm{{}^{239}Pu}$ & 0.30 & $211.12\pm0.34$     \\
			$\mathrm{{}^{241}Pu}$ & 0.05 & $214.26\pm0.33$     \\\hline\hline
		\end{tabular}
		\caption{Fission fraction and average released energy of each fissile isotope.}
		\label{tab:EnergyReleased}
		\vspace*{-0.5cm}
	\end{center}
\end{table}

\begin{figure}[!h]
	\centering
	\includegraphics[width=1.\columnwidth]{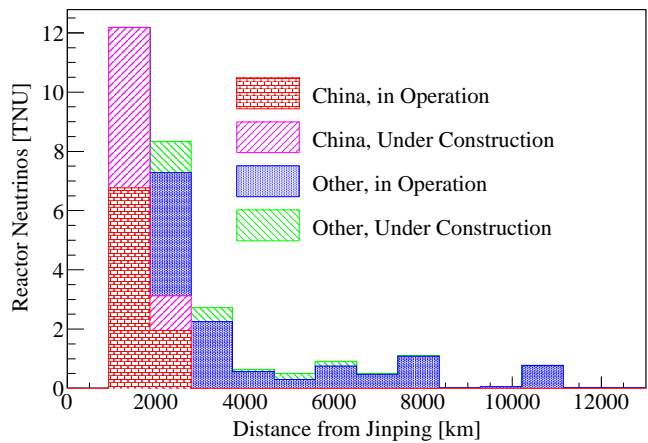}
	\caption[Flux v.s. Distance]{Reactor neutrino contribution from reactors all around the world. Data are from IAEA~\cite{IAEA:2015}.}
	\label{fig:distflux}
\end{figure}

The reactor $\bar\nu_e$ backgrounds are detected via the IBD interaction. 
The event rates for the different energy ranges are shown in Table \ref{tab:reactor}, where Full Energy Range (FER) represents $[1.8, 10.0]$ MeV and Signal Energy Range (SER) $[1.8, 3.3]$ MeV.
Figure~\ref{fig:distflux} shows the reactor neutrino contribution in FER from reactors all around the world.
Uncertainties introduced by the vacuum oscillation effect are estimated, as they are the main source of uncertainty for reactor neutrinos~\cite{Baldoncini:2015}.
Using the same technique as explained in the geoneutrino oscillation analysis and parameter uncertainty listed in Table \ref{tab:Uncer}, the uncertainty in reactor neutrino flux prediction is estimated to be 1.5\%.
The MSW effect contributes 0.5\% deviation from the central value.
This deviation is not included in the uncertainties in Table~\ref{tab:reactor}.
%Assuming an exposure of 3 kilotons $\times$ 1,500 days, the reactor neutrino background in the SER is estimated to be $60.4\pm0.7$.

\begin{table}[!h]
	\centering
	\begin{tabular}{cccccc} \hline\hline
		Event rate & \multicolumn{2}{c}{Constructed} & \multicolumn{2}{c}{Under construction} & Total \\
		(TNU)	& China & Others & China & Others &       \\\hline
		FER      & $8.9\pm0.2$  & $10.6\pm0.1$   & $6.3\pm0.1$  & $2.0\pm0.0$   & $27.8\pm0.4$  \\
		SER		  & $2.4\pm0.1$ & $2.3\pm0.0$ & $1.5\pm0.0$ & $0.6\pm0.0$ & $6.8\pm0.1$ \\\hline\hline
		%w/ load factor
%		FER      & $7.0\pm0.1$  & $8.4\pm0.1$   & $5.0\pm0.1$  & $1.6\pm0.0$   & $22.0\pm0.3$  \\
%		SER		  & $1.9\pm0.0$ & $1.8\pm0.0$ & $1.2\pm0.0$ & $0.4\pm0.0$ & $5.3\pm0.1$ \\\hline\hline
	\end{tabular}
	\caption{Reactor neutrino event rate at Jinping.}
	\label{tab:reactor}
\end{table}

\subsection{Other Backgrounds}

%Solar neutrinos also come as a possible background for the detection of geoneutrinos, with the elastic scattering process:
%\begin{equation}
%	\nu+e\to \nu +e.
%\end{equation}
%This process has relatively low cross section compared to the IBD cross section, and the electron will not coincide with a neutron to form a correlated pair of signals, thus it is unable to produce a background for the IBD sample.
Except the main background of reactor neutrinos, there are several other backgrounds applicable to geoneutrino detection.
When cosmogenic muons pass through the detector, the possibly induced $^{9}\text{Li}$-$^{8}\text{He}$ isotopes can decay to produce correlated electron and neutron signal, and thus mimic IBD events. 
Fast neutrons are also produced by cosmogenic muons near the detector.
The muon rate at Jinping is as low as $(2.0\pm0.4)\times10^{-10}/(\text{cm}^2\cdot\text{s})$, with the 6,720 meter water equivalent overburden, greatly suppressing these backgrounds~\cite{JinpingNE:2016}.
A muon veto window of 2 s is assumed, resulting in 1.1\% live time loss. 
The estimation of  $^{9}\text{Li}$-$^{8}\text{He}$  background is $(0.02\pm0.01)$/3 kilotons/1,500 days.  
The fast neutron background is estimated to be $\le0.04$/3 kilotons/1,500 days.

The $\alpha$ particles produced in the decay series of radioactive isotopes can trigger ($\alpha$, n) reactions in the liquid scintillator.
$^{210}$Po background plays a decisive role in determining the background rate of ($\alpha$, n) for geoneutrino detection.
Assuming the same level of $^{210}$Po background as Borexino~\cite{BX:2010}, $(1.7\pm0.1)$/3 kilotons/1,500 days is estimated for Jinping.
Accidental coincidence background is negligible, assuming the cleanness of the neutrino detector and fiducial volume cut to reject the natural radioactivity outside~\cite{JinpingNE:2016}.

The energy range of solar neutrinos also overlaps with that of geoneutrinos.
However, solar neutrinos do not interact through the IBD channel, and do not produce a neutron to form a correlated pair mimicking IBD signals.

In the following study, we ignore the above backgrounds summing up to $1.8/$3 kilotons/1,500 days in SER, compared to the reactor neutrino background of $(60.4\pm0.9)$/3 kilotons/1,500 days.

%%----------------------------------------------------------------------------------------
\section{Sensitivity Study for Future Experiment at Jinping}

In this section, we first present a predicted overview for IBD events at Jinping including the geoneutrino signal and reactor neutrino background.
A discussion of the geoneutrino flux measurement sensitivity and the determination of the Th/U ratio are presented.
Finally, the potential of Jinping on the test of the present BSE model is discussed.

\subsection{Predicted IBD Spectrum at Jinping}

\begin{figure}[!h]
	\centering
	\includegraphics[width=1.\columnwidth]{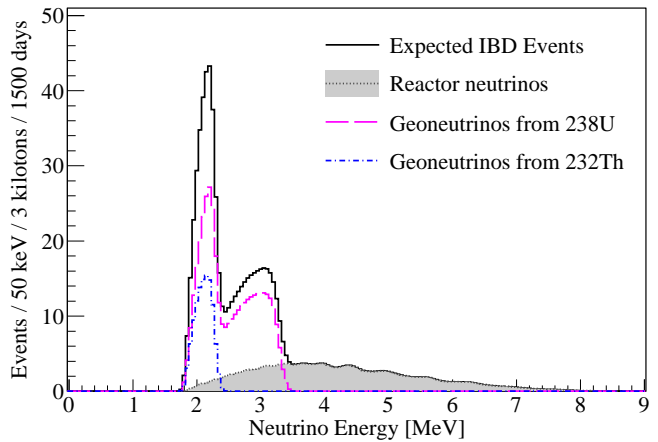}
	\caption[Event Spectrum]{Predicted IBD events at the Jinping site. $\bar\nu_e$ sources include $^{238}$U decay (magenta dashed), $^{232}$Th decay (blue dotted dashed), and man-made reactor background (gray filled). The black solid line sums up all.}
	\label{fig:eevent}
\end{figure}

With an exposure of 3 kilotons $\times$ 1,500 days, an expected IBD spectrum is obtained as shown in Fig. \ref{fig:eevent}, with a 500 p.e./MeV energy resolution assumption ($4.4\%/\sqrt {E_{vis}}$) and $50$ keV binning.
Predictions for geoneutrinos coming from the Earth crust and mantle are classified in Table~\ref{tab:CMRate}.
The event rates of geoneutrino signals and reactor neutrino background are summarized in Table~\ref{tab:CanRate}.

\begin{table}[!h]
	\centering
	\begin{tabular}{cccc}   \hline\hline
		Geo $\bar\nu_e$ (TNU)	& Crust & Mantle & BSE\\\hline
		Th   & $10.6\pm0.8$  & $2.1\pm0.5$   &  $12.7\pm1.0$  \\
		U    & $38.4\pm6.6$  & $8.3\pm2.3$   &  $46.7\pm6.7$  \\
		Th+U & $49.0\pm7.3$  & $10.4\pm2.7$  &  $59.4\pm7.6$  \\\hline\hline
	\end{tabular}
	\caption{Summary of predicted geoneutrino event rates in TNU at Jinping.}
	\label{tab:CMRate}
\end{table}

\begin{table}[!h]
	\centering
	\begin{tabular}{cccccc}   \hline\hline
		& \multicolumn{3}{c}{Geoneutrino} & \multicolumn{2}{c}{Reactor} \\
		& ~$^{238}$U~ & ~$^{232}$Th~  & ~Total~   &     ~FER~ & ~SER~   \\\hline
		Event Rate ( TNU )     &  ~46.7~    & ~12.7~     & ~59.4~   &  ~27.8~  &  ~6.8~ \\
		Total Events     &  ~414.5~    & ~113.6~     & ~527.3~   &  ~246.8~  &  ~60.4~ \\\hline\hline
		%LF=0.8
%		Event Rate ( TNU )     &  ~46.7~    & ~12.7~     & ~59.4~   &  ~22.0~  &  ~5.3~ \\
%		Total Events     &  ~414.5~    & ~113.6~     & ~527.3~   &  ~194.4~  &  ~47.0~ \\\hline\hline
	\end{tabular}
	\caption{Geoneutrino and reactor neutrino event rates and total events with an exposure of 3 kilotons $\times$ 1,500 days at Jinping.}
	\label{tab:CanRate}
\end{table}

\subsection{Sensitivity for Geoneutrinos}

To obtain the sensitivity for geoneutrinos at Jinping, a toy Monte Carlo with an exposure of 3 kilotons $\times$ 1,500 days is employed.
The simulated spectrum with signal and background is randomly sampled according to the exposure and fitted using the maximum likelihood method in the energy range $E_v\in[1.8, 6.8]$ MeV.
The fitting function is
\begin{equation}
	\begin{aligned}
		N(E)=&NR_{geo}\left[R_\text{U}\tilde f_\text{U}(E)+\left(1-R_\text{U}\right)\tilde f_\text{Th}(E)\right]\\
		&+N(1-R_{geo})\tilde f_\text{R}(E),
	\end{aligned}
\end{equation}
where $N(E)$, $R_{geo}$, and $R_\text{U}$ are the free fit parameters. $N(E)$ is the number of events observed in the energy bin $E$, and $R_{geo}$ and $R_\text{U}$ denote the fraction of the number of geoneutrino events in the IBD events and the fraction of $^{238}$U geoneutrino events in the total geoneutrino events.
$\tilde f(E)$'s are the normalized oscillated electron antineutrino energy spectra at Jinping weighted by the IBD cross section. 
The subscripts U, Th, and R denote $^{238}$U, $^{232}$Th, and reactors. 
The process of sampling and fitting is repeated 10,000 times.

The total geoneutrino event rate can be calculated as 
\begin{equation}
	N_{geo}=N\cdot R_{geo},
\end{equation}
and the Th/U IBD event ratio locally measured in Jinping is 
\begin{equation}
	R(\text{Th}/\text{U})_\text{IBD}=(1-R_\text{U})/R_\text{U}.
\end{equation}
The Th/U mass ratio in BSE is then expressed as,
\begin{equation}
	\begin{aligned}
		R(\text{Th}/\text{U})_{m}=&
		R(\text{Th}/\text{U})_\text{IBD}\cdot
		\frac{\langle P^{U}_{ee}\rangle^{'}\tilde\sigma_\text{U}}{\langle P^\text{Th}_{ee}\rangle^{'}\tilde\sigma_\text{Th}}\\
		&\cdot \frac{X_\text{U}\lambda_\text{U}n_\nu(\text{U})\mu_\text{Th}}{X_\text{Th}\lambda_\text{Th}n_\nu(\text{Th})\mu_\text{U}}.
	\end{aligned}
\end{equation}
The notations are the same as in Eqs.~(\ref{Eq:fluxE}) and~(\ref{Eq:flux}) and Table~\ref{tab:Pee}, $\tilde\sigma$'s are the effective cross section for HPE $i$, and $\tilde\sigma_i=\int \sigma(E)f_i(E)dE/\int f_i(E)dE$.
The predicted central value for the Th/U ratio at Jinping is $R(\text{Th}/\text{U})_\text{IBD}=0.27$ or $R(\text{Th}/\text{U})_{m}=4.1$.

Under the assumption of a precisely known reactor neutrino spectrum and a free reactor neutrino rate, the precisions of the geoneutrino measurements at Jinping can be concluded as shown in Table~\ref{tab:Sensitivity}.
\begin{table}[!h]
	\centering
	\begin{tabular}{cc}   \hline\hline
		Measurement & Precision (\%) \\\hline
		$N_{geo}$& 4.6\% \\
		$R(\text{Th}/\text{U})_\text{IBD}$ & 26.3\%\\\hline\hline
	\end{tabular}
	\caption{Precisions of the geoneutrino measurements at Jinping.}
	\label{tab:Sensitivity}
\end{table}

\subsection{BSE Model Test}

\begin{figure}[!h]
	\centering
	\includegraphics[width=1.\columnwidth]{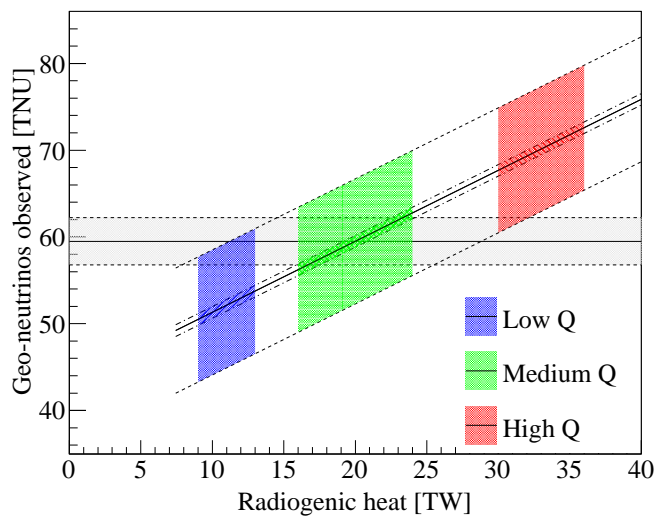}
	\caption[Earth Model]{The slanted solid line represents the predicted observed geoneutrino from the Earth crust and mantle, vs radiogenic heat from HPEs, while the slanted dashed lines represent the uncertainty in prediction. Colored bands represent different BSE models, blue for low-Q model, green for medium-Q, and red for high-Q estimation. Dashed lines represent the present uncertainty in crustal neutrino flux prediction. The gray band corresponds to the prediction of observed geoneutrino signals.}
	\label{fig:EarthModel}
\end{figure}

Several BSE models can be tested with geoneutrinos coming from the mantle, as shown in Fig. \ref{fig:EarthModel}.
The upper and lower dashed lines incorporate the uncertainty in the crustal contribution prediction.
%, including the 15.0\% from the crustal Earth model and the 1.7\% from oscillation parameters.
The prediction for the experimental geoneutrino event rate from the crust and mantle (gray band) is compared with the expectations for the different BSE models from the low-Q~\cite{Javoy:2010}, medium-Q~\cite{ArevaloJr:2010}, and high-Q~\cite{turcotte:1982} estimates (color bands), with the central value calculated with medium-Q model.
The sloped band indicates the response between the isotropic mantle radiogenic heat (assuming a fixed Th/U and K/Th ratio) and $\bar\nu_e$ flux from the mantle, and its starting point of 7.4 TW and 49.0 TNU corresponds to the contribution from crustal HPEs.
The vertical width of the band represents uncertainty in the present crustal neutrino flux prediction, which is crucial to the BSE model test.

As discussed in Sec. III, the uncertainty in the geoneutrino prediction comes from the Earth model and the oscillation parameters.
At present, the Earth model contributes $\pm$15.0\% uncertainty for crustal geoneutrino prediction, while oscillation parameters contribute $^{+1.8\%}_{-1.7\%}$.
Nearly 50\% of the geoneutrino signals comes within 300 km distance from the detector~\cite{Sramek:2016}; therefore, a clear understanding of the local geological environment is fundamentally important.
The local geology around Jinping has been heavily studied because of the many devastating earthquakes in the region, and still requires further effort toward an accurate local lithospheric model.
With improvements on local crust composition and oscillation parameters, supposing an optimistic $\pm$1.0\% uncertainty in the Earth model, and $\pm$1.0\% uncertainty in oscillation parameters, this test on BSE models will be significantly improved as shown by the dashed dotted lines in Fig. \ref{fig:EarthModel}.

\section{Earth Core Fission Neutrinos}

%In this study, the Earth core fission reactors, or georeactors, are assumed to be uniformly distributed inside the inner core of Earth.
The Earth core fission reactors, or georeactors, are supposed to be fission reactors inside the inner core of the Earth. 
The impact of georeactors' distribution is ignorable.
Simulation shows that to sustain such a long-term self-burning georeactor, several conditions must be satisfied.
The thermal power should be within the range of 3-30 TW, and the fission fuel is composed of 74.6\% $^{235}$U and 24.9\% $^{238}$U~\cite{Herndon:2005}.
This yields an electron antineutrino spectrum very similar to reactor neutrino spectra.

Current experimental results set an upper limit of 4.5 TW for the georeactor from Borexino~\cite{BX:2013} and 3.7 TW by KamLAND~\cite{KL:2013} with 95\% C.L..

To derive an upper limit at Jinping, the CLs+b method was applied~\cite{CLS:2002}.
Assuming a total exposure of 3 kilotons x 300 days and a known reactor neutrino rate, an upper limit of 1.4 TW was obtained at 95\% C.L., compared to the 3 TW theoretical lower limit of georeactor power.

\section{Conclusion}

We discussed in this paper the potential of a 3-kiloton scintillation detector at Jinping Observatory to detect geoneutrinos and to test different Earth models.
The expected geoneutrino signals and background rates are reported, $S_\text{U}=46.7\pm6.7$ TNU, $S_\text{Th}=12.7\pm1.0$ TNU, and $S_\text{R}=27.8\pm0.4$ TNU in the full energy range (FER) and $S_\text{R}=6.8\pm0.1$ TNU in the signal energy range (SER) $[1.8, 3.3]$ MeV.

An analysis on the uncertainty from the oscillation parameters was performed, and an intrinsic $^{+1.8\%}_{-1.7\%}$ uncertainty is presented. 
This is smaller than the $\pm$15.0\% uncertainty in the present crust model prediction.
The MSW oscillation yields a $+$0.3\% deviation from the predicted flux of geoneutrinos based on the vacuum oscillation.

For an exposure of 3 kilotons $\times$ 1,500 days, the geoneutrino flux dominated by the crustal contribution can be measured with a precision of 4.6\% with a free Th/U ratio, and the ratio itself can be measured with a precision of 26.3\%, thus greatly enhancing the global effort in discriminating between different BSE models.
The proposed 3-30 TW Earth core fission reactor can be confirmed or excluded within 300 days of live time.

\section*{Acknowledgments}

We wish to thank Professor William F. McDonough and Dr. Yufei Xi for the valuable discussions.
Special thanks go to Dr. Ondrej Sramek for sharing his geoneutrino calculation program and the kind suggestions on this paper.
This work is supported in part by, the National Natural Science Foundation of China (Grants No. 11235006 and No. 11475093), the Tsinghua University Initiative Scientific Research Program (Grant No. 20121088035), the Key Laboratory of Particle and Radiation Imaging (Tsinghua University), and the CAS Center for Excellence in Particle Physics. 

\nocite{*}

\bibliography{apssamp}% Produces the bibliography via BibTeX.

\end{document}